\documentclass{article}
\usepackage{spconf,amsmath,graphicx}

\usepackage{spconf,amsmath,graphicx,spconf,epsfig,stfloats,hyperref,multirow,booktabs,caption,color,cite,amssymb}
\usepackage{subcaption}

\title{FreeVC: Towards high-quality text-free one-shot voice conversion}
\name{Jingyi Li$^{1,2}$      Weiping Tu$^{1,2,*}$   Li Xiao$^{1,2}$    \thanks{$*$ Corresponding author.}}

\address{$^{1}$National Engineering Research Center for Multimedia Software, School of Computer Science, \\Wuhan University, Wuhan 430072, China\\
$^{2}$Hubei Key Laboratory of Multimedia and Network Communication Engineering, \\Wuhan University,Wuhan 430072, China}
\begin{document}
%
\maketitle
\begin{abstract}
Voice conversion (VC) can be achieved by first extracting source content information and target speaker information, and then reconstructing waveform with these information. However, current approaches normally either extract dirty content information with speaker information leaked in, or demand a large amount of annotated data for training. Besides, the quality of reconstructed waveform can be degraded by the mismatch between conversion model and vocoder. In this paper, we adopt the end-to-end framework of VITS for high-quality waveform reconstruction, and propose strategies for clean content information extraction without text annotation. We disentangle content information by imposing an information bottleneck to WavLM features, and propose the spectrogram-resize based data augmentation to improve the purity of extracted content information. Experimental results show that the proposed method outperforms the latest VC models trained with annotated data and has greater robustness.
\end{abstract}
\section{Introduction}
\label{sec:intro}
Voice conversion (VC) is a technique that alters the voice of a source speaker to a target style, such as speaker identity~\cite{spki}, prosody~\cite{prosody} and emotion~\cite{emotion}, while keeping the linguistic content unchanged. In this paper, we focus on the speaker identity conversion under one-shot setting, i.e., given only one utterance of target speaker as reference.

A typical approach of one-shot voice conversion is to disentangle content information and speaker information from source and target speech, respectively, and then use them to reconstruct the converted speech~\cite{autovc}. As a result, the quality of converted speech relys on (1) the disentanglement ability of VC model, and (2) the reconstruction ability of VC model.

Based on how a VC system disentangles content information, we can categorize current VC approaches into text-based VC and text-free VC. A popular text-based VC approach is to use an automatic speech recognition (ASR) model to extract phonetic posteriorgram (PPG) as content representation~\cite{ppg}\cite{ppgvc}. Some researchers have also resolved to leveraging shared linguistic knowledge from a text-to-speech (TTS) model~\cite{ttsvc}\cite{cotatron}. However, these approaches require an extensive amount of annotated data for training the ASR or TTS model. Data annotation is costly, and the accuracy and granularity, e.g. phoneme level and grapheme level, of annotation affects the model performance. To avoid the concerns of text-based approaches, text-free approaches that learn to extract content information without the guidance of text annotation have been explored. Typical text-free approaches include information bottleneck~\cite{autovc}, vector quantization~\cite{vqvc+}, instance normalization~\cite{againvc}, etc. However, their performance generally lags behind text-based approaches~\cite{vcc2020}. This can be attributed to the fact that the content information they extract is more easily to have source speaker information leaked in.

Many VC systems adopt a two-stage reconstruction pipe-line~\cite{ppgvc}\cite{autovc}. A conversion model converts the source acoustic features into target speaker’s voice in the first stage, while a vocoder transforms the converted features into waveform in the second stage. The two models are usually trained separately. However, the acoustic feature predicted by conversion model has a different distribution from that the vocoder uses during training, which is from the real speech. This feature mismatch problem, which also exists in TTS, can degrade the quality of reconstructed waveform~\cite{mismatch}. VITS~\cite{vits} is a one-stage model that can do both TTS and VC. By connecting models of the two stages through latent variables of a conditional variational autoencoder (CVAE), the feature mismatch is reduced. By adopting adversarial training, the quality of reconstructed waveform is further improved. However, VITS is a text-based model and is limited to many-to-many VC, i.e. the source and target speakers are all seen speakers.

In this paper, we propose a text-free one-shot VC system named FreeVC, which adopts the framework of VITS for its brilliant reconstruction ability, but learns to disentangle content information without the need of text annotation. The recent success of speech self-supervised learning (SSL) in downstream tasks such as speech recognition~\cite{xlsr53}, speaker verification~\cite{sslasv} and voice conversion~\cite{s3prlvc} has demonstrated the potential power of SSL features over traditional acoustic features like mel-spectrograms. We use WavLM~\cite{wavlm} to extract SSL features from waveform, and introduce a bottleneck extractor to extract content information from SSL features. We also propose spectrogram-resize (SR) based data augmentation, which distorts speaker information without changing content information, to strengthen the disentanglement ability of the model. To achieve one-shot VC, we use a speaker encoder for speaker information extraction. Our code~\footnote{\url{https://github.com/OlaWod/FreeVC}} and demo page~\footnote{\url{https://olawod.github.io/FreeVC-demo}} are publicly available.

\section{Methods}
\label{sec:architecture}
\begin{figure*}
\centering
\begin{subfigure}{\columnwidth}
  \includegraphics[width=\textwidth]{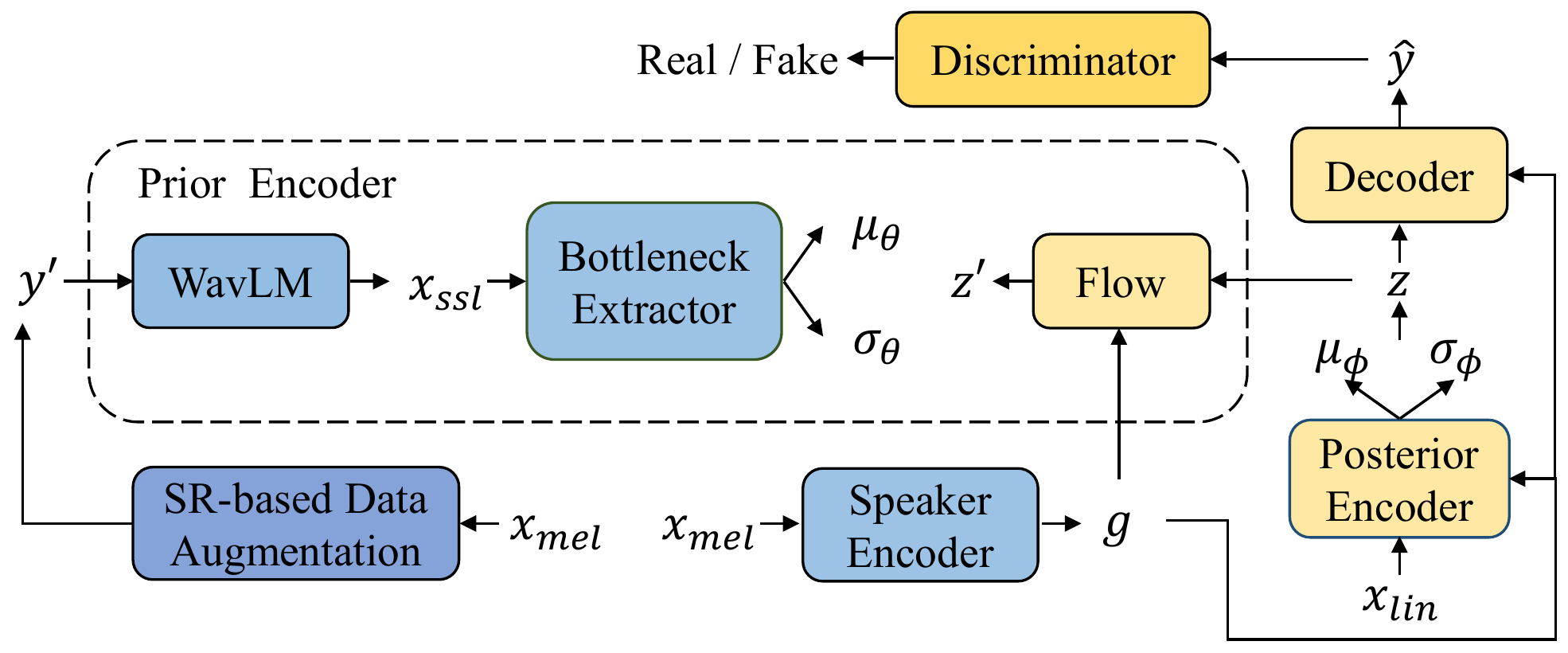}
  \caption{Training procedure}
  \label{fig:aug1}
\end{subfigure}%
\hfill
\begin{subfigure}{\columnwidth}
  \includegraphics[width=\textwidth]{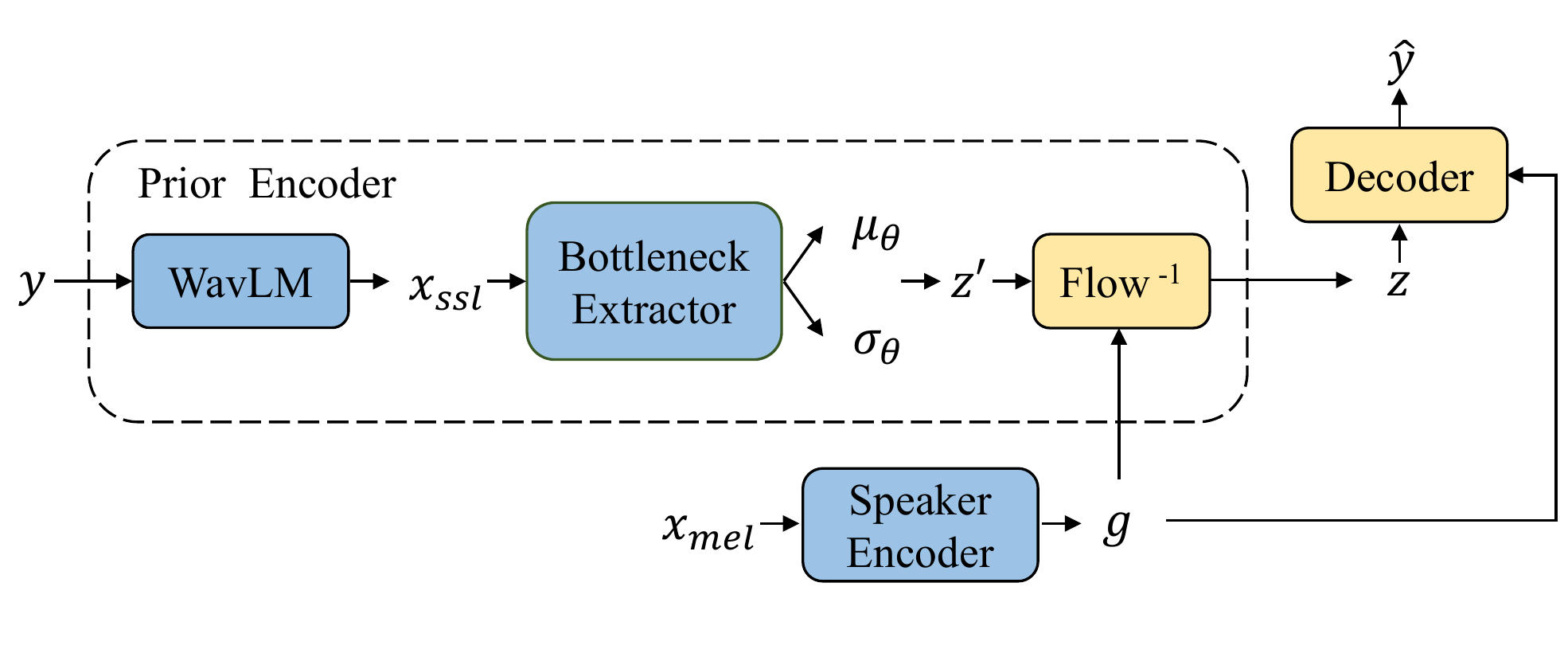}
  \caption{Inference procedure}
  \label{fig:aug2}
\end{subfigure}

\caption{Training and inference procedure of FreeVC. Here $y$ denotes source waveform, $y’$ denotes augmented waveform, $\hat{y}$ denotes converted waveform, $x_{mel}$ denotes mel-spectrogram, $x_{lin}$ denotes linear spectrogram, $x_{ssl}$ denotes SSL feature, and $g$ denotes speaker embedding.}
\label{fig:model}
\vspace{-1em}
\end{figure*}

As illustrated in Fig.\ref{fig:model}, the backbone of FreeVC is inherited from VITS, which is a CVAE augmented with GAN training. Different from VITS, the prior encoder of FreeVC takes raw waveform as input instead of text annotation, and has a different structure. The speaker embedding is extracted by a speaker encoder to perform one-shot VC. In addition, FreeVC adopts a different training strategy and inference procedure. We will present the details in the following subsections.


\subsection{Model Architecture}
\label{sssec:bottleneck}

FreeVC contains a prior encoder, a posterior encoder, a decoder, a discriminator and a speaker encoder, where the architecture of the posterior encoder, decoder and discriminator follows VITS. We will focus on describing the prior encoder and speaker encoder in the following.

\subsubsection{Prior Encoder}
\label{sssec:bottleneck}

The prior encoder contains a WavLM model, a bottleneck extractor and a normalizing flow. The WavLM model and bottleneck extractor are in charge of extracting content information in the form of modeling distribution $N(z'; \mu_{\theta}, \sigma_{\theta}^2)$. The WavLM model takes raw waveform as input and produces 1024-dimensional SSL feature $x_{ssl}$ containing both content information and speaker information. In order to remove the unwanted speaker information contained in $x_{ssl}$, the 1024-dim $x_{ssl}$ is put into the bottleneck extractor and converted into $d$-dim representation, where $d$ is much smaller than 1024. This huge dimension gap imposes an information bottleneck, forcing the resulted low-dimentional representation to discard content-irrelevant information like noise or speaker information. Next, the $d$-dim hidden representation is projected into $2d$-dim hidden representation, which is latter split into $d$-dim $\mu_{\theta}$ and $d$-dim $\sigma_{\theta}$.  The normalizing flow, which conditions on speaker embedding $g$, is adopted to improve the complexity of prior distribution. Following VITS, it is composed of multiple affine coupling layers~\cite{realnvp} and is made to be volume-preserving with the Jacobian determinant $|det\frac{\partial z'}{\partial z}|$ of 1.

\subsubsection{Speaker Encoder}
\label{sssec:bottleneck}

We use two types of speaker encoder: pretrained speaker encoder and non-pretrained speaker encoder. The pretrained speaker encoder is a speaker verification model trained on datasets with large amounts of speakers. It is widely used in VC and is considered to be superior to non-pretrained speaker encoder. We adopt the one employed in ~\cite{ppgvc}. The non-pretrained speaker encoder is jointly trained with the rest of the model from scratch. We use a simple LSTM-based architecture and believe that if the extracted content representation is clean enough, the speaker encoder will learn to model the missing speaker information.

\subsection{Training Strategy}
\label{sssec:sr}

\begin{figure}
\begin{subfigure}{\columnwidth}
  \centering
  \includegraphics[width=\columnwidth]{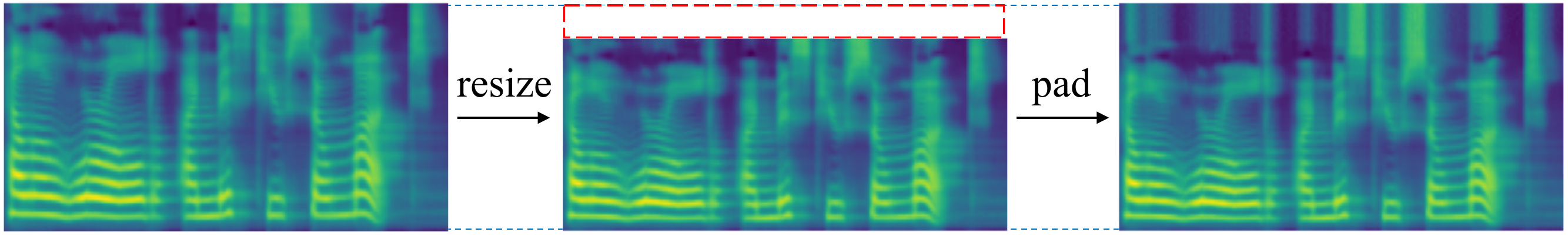}
  \caption{Resize ratio $r<1$}
  \label{fig:aug1}
\end{subfigure}%
\\
\begin{subfigure}{\columnwidth}
  \centering
  \includegraphics[width=\columnwidth]{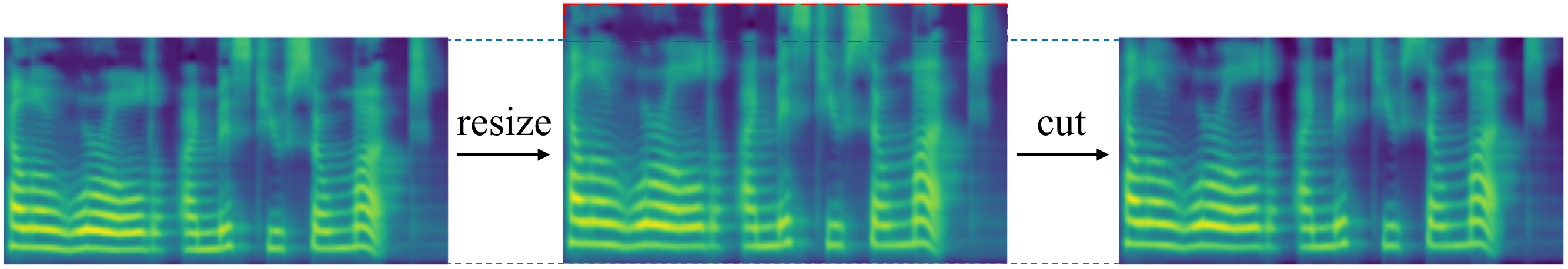}
  \caption{Resize ratio $r>1$}
  \label{fig:aug2}
\end{subfigure}
\caption{Vertical spectrogram-resize operation.}
\label{fig:vsr}
\vspace{-1em}
\end{figure}

\subsubsection{SR-based Data Augmentation}
\label{sssec:sr}
A too narrow bottleneck will lose some content information, while a too wide bottleneck will contain some speaker information~\cite{autovc}. Instead of carefully tuning the bottleneck size, we resolve to SR-based data augmentation to help the model to learn to extract clean content information by distorting speaker information in the source waveform. Unlike works~\cite{speechsplit2}\cite{nansy} that use various signal processing techniques to corrupt speaker information, our method is much easier to implement, and does not require complicated signal processing knowledge. 

Our proposed SR-based data augmentation includes three steps: (1) get mel-spectrogram $x_{mel}$ from waveform $y$; (2) conduct vertical SR operation to $x_{mel}$, resulting in modified mel-spectrogram $x_{mel}'$; (3) reconstruct waveform $y'$ from $x_{mel}'$ with a neural vocoder. The vertical SR operation is depicted in Fig.\ref{fig:vsr}. A mel-spectrogram can be seen as an image with horizontal time axis and vertical frequency bin axis. Vertical SR operation first resizes mel-spectrogram to a certain ratio $r$ vertically using bilinear interpolation, and then pads or cuts the resized mel-spectrogram to the original shape. If the ratio $r$ is less than $1$, we pad the squeezed mel-spectrogram at the top with the sum of highest frequency bin value and Gaussian noise, resulting in speech with lower pitch and closer formant distance; else, we cut redundant frequency bins at the top of the stretched mel-spectrogram, resulting in speech with higher pitch and farther formant distance. By training with augmented speech, the model will better learn to extract the unchanged content information shared in each ratio $r$. In addition to vertical SR, we can also use horizontal SR to produce time-scale modified waveforms.

\subsubsection{Training Loss}
\label{ssec:backbone}
The training loss is divided into  CVAE-related loss and GAN-related loss. The CVAE-related loss consists of reconstruction loss $L_{rec}$, which is the $L_1$ distance between target and predicted mel-spectrogram, and KL loss $L_{kl}$, which is the KL divergence between prior distribution $p_{\theta}(z|c)$ and posterior distribution $q_{\phi}(z|x_{lin})$, where
\begin{align}
&q_{\phi}(z|x_{lin}) = N(z; \mu_{\phi}, \sigma_{\phi}^2),\\
&p_{\theta}(z|c) = N(z'; \mu_{\theta}, \sigma_{\theta}^2)|det\frac{\partial z'}{\partial z}|.
\end{align}
Here the condition $c$ is content information contained in waveform $y/y’$. By minimizing $L_{kl}$, the feature mismatch problem can be reduced. The GAN-related loss consists of adversarial loss~\cite{least} $L_{adv}(D)$ and $L_{adv}(G)$ for discriminator $D$ and generator $G$, and feature matching loss~\cite{fm} $L_{fm}(G)$ for generator $G$. Finally, the training loss of FreeVC can be expressed as:
\begin{align}
    L(G) &= L_{rec} + L_{kl} + L_{adv}(G) + L_{fm}(G),\\
    L(D) &= L_{adv}(D).
\end{align}

\subsection{Inference Procedure}
\label{sssec:sr}

Different from VITS, which extracts content information through posterior encoder and normalizing flow in prior encoder during VC inference, FreeVC extracts content information through WavLM and bottleneck extractor in prior encoder during inference as with in training. Such that the extracted content representation will not be affected by the quality of source speaker embedding.

\section{Experiments}
\label{sec:exp}

\begin{table*}[]
\centering
\caption{Subjective evaluation results in terms of 5-scale MOS and SMOS with 95\% confidence intervals under seen-to-seen, unseen-to-seen and unseen-to-unseen scenarios. For reference, we also report scores of source utterances.}
\label{tab:sbj}
\resizebox{\textwidth}{!}{%
\begin{tabular}{l|cc|cc|cc}
\hline
                & \multicolumn{2}{c|}{seen-to-seen}                                           & \multicolumn{2}{c|}{unseen-to-seen}                                         & \multicolumn{2}{c}{unseen-to-unseen}                                        \\ \hline
                & MOS                                  & SMOS                                 & MOS                                  & SMOS                                 & MOS                                  & SMOS                                 \\ \hline
VQMIVC          & $\quad$2.31$\pm$0.09$\quad$          & $\quad$2.10$\pm$0.08$\quad$          & $\quad$1.50$\pm$0.08$\quad$          & $\quad$1.71$\pm$0.08$\quad$          & $\quad$1.49$\pm$0.08$\quad$          & $\quad$1.29$\pm$0.05$\quad$          \\
BNE-PPG-VC      & $\quad$2.80$\pm$0.12$\quad$          & $\quad$2.95$\pm$0.12$\quad$          & $\quad$2.89$\pm$0.10$\quad$          & $\quad$2.83$\pm$0.10$\quad$          & $\quad$3.44$\pm$0.08$\quad$          & $\quad$2.63$\pm$0.10$\quad$          \\
YourTTS         & $\quad$3.46$\pm$0.10$\quad$          & $\quad$3.25$\pm$0.09$\quad$          & $\quad$2.54$\pm$0.10$\quad$          & $\quad$2.50$\pm$0.10$\quad$          & $\quad$2.87$\pm$0.09$\quad$          & $\quad$1.97$\pm$0.09$\quad$          \\ \hline
FreeVC          & $\quad$3.99$\pm$0.09$\quad$          & \textbf{$\quad$3.80$\pm$0.09$\quad$} & $\quad$4.06$\pm$0.08$\quad$          & \textbf{$\quad$3.77$\pm$0.09$\quad$} & \textbf{$\quad$4.06$\pm$0.08$\quad$} & \textbf{$\quad$2.83$\pm$0.08$\quad$} \\
FreeVC (w/o SR) & $\quad$3.85$\pm$0.10$\quad$          & $\quad$3.50$\pm$0.10$\quad$          & $\quad$3.88$\pm$0.08$\quad$          & $\quad$3.58$\pm$0.08$\quad$          & $\quad$3.97$\pm$0.09$\quad$          & $\quad$2.80$\pm$0.09$\quad$          \\
FreeVC-s        & \textbf{$\quad$4.01$\pm$0.09$\quad$} & $\quad$3.75$\pm$0.09$\quad$          & \textbf{$\quad$4.08$\pm$0.08$\quad$} & $\quad$3.68$\pm$0.09$\quad$          & $\quad$4.02$\pm$0.09$\quad$          & $\quad$2.78$\pm$0.09$\quad$          \\ \hline
Source          & $\quad$4.32$\pm$0.08$\quad$          & -                                    & $\quad$4.11$\pm$0.10$\quad$          & -                                    & $\quad$4.17$\pm$0.09$\quad$          & -                                    \\ \hline
\end{tabular}%
}
\end{table*}

\subsection{Experimental Setups}
\label{ssec:datapre}
We conduct experiments on VCTK~\cite{vctk} and LibriTTS~\cite{libritts}. Only VCTK corpus is used for training. For VCTK, we use data from 107 speakers, in which 314 utterances  (2 sentences per speaker) are randomly selected for validation, 10700 utterances (10 sentences per speaker) for test, and the rest for training. For LibriTTS, we use the test-clean subset for test. 

All audio samples are downsampled to 16 kHz. Linear spectrograms and 80-band mel-spectrograms are calculated using  short-time Fourier transform. The FFT, window, and hop size are set to 1280, 1280, and 320, respectively. We set the dimension $d$ of bottleneck extractor to 192. For the SR-based data augmentation, the resize ratio $r$ ranges from 0.85 to 1.15. And a HiFi-GAN v1 vocoder~\cite{hifi} is used to transform the modified mel-spectrogram into waveform. Our models are trained up to 900k steps on a single NVIDIA 3090 GPU. The batch size is set to 64 with a maximum segment length of 128 frames.

Three baseline models are selected to be compared with the proposed method: (1) VQMIVC~\cite{vqmivc}, a text-free model that uses non-pretrained speaker encoder; (2) BNE-PPG-VC~\cite{ppgvc}, a text-based model that uses pretrained speaker encoder; (3) YourTTS~\cite{yourtts}, a text-based model that extends VITS to one-shot setting by introducing a pretrained speaker encoder. Three versions of the proposed method are tested: (1) FreeVC-s, the proposed model that uses non-pretrained speaker encoder. (2) FreeVC, the proposed model that uses pretrained speaker encoder. (3) FreeVC (w/o SR), the proposed model that uses pretrained speaker encoder, but is trained without SR-based data augmentation.

\subsection{Evaluation Metrics}
\label{sec:result}
We conduct evaluations both subjectively and objectively. For subjective evaluation, 15 participants are invited to evaluate the naturalness and speaker similarity of the speech in terms of 5-scale mean opinion score (MOS) and similarity mean opinion score (SMOS), respectively. We randomly select 6 seen speakers (3 male, 3 female) from VCTK, 6 unseen speakers (3 male, 3 female) from LibriTTS, and conduct evaluation in seen-to-seen, unseen-to-seen, and unseen-to-unseen scenarios separately. For objective evaluation, we use three metrics: WER, CER and F0-PCC. Word error rate (WER) and character error rate (CER) between source and converted speech are obtained by an ASR model~\footnote{\href{https://huggingface.co/facebook/hubert-large-ls960-ft}{https://huggingface.co/facebook/hubert-large-ls960-ft}}. F0-PCC is the Pearson correlation coefficient~\cite{pcc} between F0 of source and converted speech. We randomly select 400 utterances (200 from VCTK, 200 from LibriTTS) as source speech, and 12 speakers (6 seen, 6 unseen) as target speaker.

\subsection{Results and Analysis}
\label{ssec:ablation}

\subsubsection{Speech Naturalness and Speaker Similarity}
\label{ssec:ablation}
The MOS and SMOS results in Table~\ref{tab:sbj} demonstrate that the proposed models outperform all baseline models in all scenarios in terms of both speech naturalness and speaker similarity. In addition, we observe that all baselines suffer from quality degradation when the quality of source speech is low, for example, with a low recording quality or an unclear pronunciation, while our proposed models are barely affected, which shows the robustness of proposed content extraction method.

Within the three versions of the proposed method, FreeVC (w/o SR) achieves lower speech naturalness and speaker similarity than FreeVC. This indicates that model trained without SR-based data augmentation has some source speaker information leaked to the bottleneck, making it harder to reconstruct satisfactory waveform. FreeVC-s has a similar performance with FreeVC, demonstrating that a pretrained speaker encoder is not a dominating factor of our method's performance, and a simple non-pretrained speaker encoder is able to match the performance with a pretrained speaker encoder. FreeVC performs better than FreeVC-s in unseen-to-unseen scenario, which indicates that a speaker encoder pretrained on large amounts of speakers can improve the performance for unseen targets. 

\subsubsection{Speech Intelligence and F0 Variation Consistency}
\label{ssec:ablation}
It can be seen in Table~\ref{tab:obj} that our proposed models achieve lower WER and CER than all baseline models, even the text-based ones. This indicates that the proposed method can preserve the linguistic content of source speech well. The F0-PCC results show that our proposed method has higher F0 variation consistency with the source speech, which demonstrates that the proposed method can effectively maintain the prosody of source speech. Besides, we observe that training with SR-based data augmentation improves both speech intelligence and F0 variation consistency slightly.

\begin{table}[]
\centering
\caption{Objective evaluation results. For WER and CER, the smaller the better. F0-PCC ranges from -1 to 1 and the higher the better.}
\label{tab:obj}
\resizebox{\columnwidth}{!}{%
\begin{tabular}{l|cc|c}
\hline
                 & WER           & CER           & F0-PCC             \\ \hline
VQMIVC     & $\quad$50.68\%$\quad$         & $\quad$29.61\%$\quad$         & $\quad$0.665$\quad$               \\
BNE-PPG-VC     & 6.54\%         & 2.50\%         & 0.718             \\
YourTTS & 12.87\%          & 5.70\%          & 0.736                \\ \hline
FreeVC            & 4.35\% & 1.53\% & \textbf{0.778}  \\
FreeVC (w/o SR)           & 4.92\%          & 1.77\%          & 0.762           \\
FreeVC-s           & \textbf{4.23\%}          & \textbf{1.46\%}          & 0.768                \\ \hline
\end{tabular}%
}
\end{table}

\section{Conclusion}
\label{sec:conclusion}
This paper proposes FreeVC, a text-free one-shot voice conversion system. We adopt the framework of VITS for high-quality waveform reconstruction. The content information is extracted from the bottleneck of WavLM features. We also propose SR-based data augmentation to improve the disentanglement ability of the model. Experimental results demonstrate the superiority of proposed methods. In the future, we will investigate speaker adaptation methods to improve similarity for unseen target speakers with little data.



\vfill\pagebreak

\bibliographystyle{IEEEbib}
\bibliography{strings}

\end{document}